# Puppets or partners?

# Governing cyborg propaganda in the digital public square

Jonas R. Kunst[1*], Kinga Bierwiaczonek[2], Meeyoung Cha[3], Omid V. Ebrahimi[4], Marc Fawcett-Atkinson[5], Asbjørn Følstad[6], Anton Gollwitzer[1], Nils Köbis[7,8], Gary Marcus[9], Jon Roozenbeek[10,11], Daniel Thilo Schroeder[6], Jay J. Van Bavel[9,12], Sander van der Linden[10], Rory White[5], & Live Leonhardsen Wilhelmsen[1]

1. BI Norwegian Business School
2. University of York
3. Max Planck Institute for Security and Privacy
4. Oxford University
5. Canada's National Observer
6. SINTEF
7. Research Center Trustworthy Data Science and Security, University Duisburg-Essen
8. Max Planck Institute for Human Development
9. New York University
10. University of Cambridge
11. Vrije Universiteit Amsterdam
12. Norwegian School of Economics

*Corresponding author (jonas.r.kunst@bi.no). All authors starting from the second author position are listed alphabetically.

**Abstract**

The distinction between genuine grassroots activism and automated influence operations is collapsing. While contemporary policy debates prioritize fully autonomous generative agents and synthetic content, this paper offers a conceptual contribution: we develop 'cyborg propaganda,' a closed-loop architecture combining verified human accounts with algorithmic automation to generate personalized content at scale, as a distinct and undertheorized threat to democratic discourse. By relying on verified citizens to ratify AI-generated messages, these campaigns exploit a regulatory gray zone that frameworks built on the human/bot binary (including the EU AI Act and Section 230) are structurally unable to address. Drawing on a conceptual analysis of coordination platforms and comparative examination of governance frameworks across democratic and non-democratic contexts, we analyze this paradox across micro, meso, and macro levels. We examine whether cyborg propaganda democratizes political power by unionizing influence or reduces citizens to cognitive proxies of a hidden directive, arguing that it shifts political discourse from a contest of ideas to a battle of algorithmic campaigns. We propose three regulatory responses: classifying coordination hubs as political action committees to enforce supply-chain transparency; mandating researcher access to platform data through DSA-style mechanisms; and establishing risk standards penalizing amplification of synthetically coordinated content. Comparative analysis reveals that viability varies structurally. Democratic states are simultaneously the most capable of regulation and the most rule-of-law constrained. By contrast, non-democratic actors face no comparable accountability, making international risk standards the primary cross-border enforcement mechanism. These measures collectively shift regulatory focus from individual speech acts toward the industrial manufacturing of political consensus.

*Keywords:* AI governance, cyborg propaganda, coordinated inauthentic behavior, generative AI policy, internet governance, synthetic media

## Introduction

A push notification lights up five thousand smartphones across the country. Not a breaking news alert, but a directive from a partisan campaigning app: *"The narrative on the tax bill is slipping. Post this now to regain control."* With two taps, users spanning the social spectrum open the app and receive unique, AI-written captions tailored to their specific background and tone. They then post these on their personal social media networks. Within minutes, the topic is trending. To observers, this mimics spontaneous yet converging public sentiment. In reality, it is a calculated, synchronized strike.

This phenomenon is rooted in documented coordination infrastructure spanning almost a decade. Act.IL, active until 2022 and analyzed by the Digital Forensic Research Lab (DFRLab, 2019), offered one of the first publicly documented platforms for organized civilian amplification of political messaging. Current commercial tools including Greenfly, SocialToaster, and GoLaxy extend and automate this model. These platforms' promotional materials, treated here as documentary evidence of stated capabilities rather than verified deployment claims, openly describe coordinated amplification architectures: Greenfly, for instance, markets the ability to "synchronize an army of advocates to amplify your message" (Greenfly, 2026). Beyond commercial promotion, two independent investigative bodies provide empirical grounding for the more advanced, AI-assisted architecture we analyze in this paper. First, leaked internal documents from the Chinese firm GoLaxy, independently reviewed by researchers at Vanderbilt University and the Doublethink Lab (Doublethink Lab, 2026; Goldstein & Benson, 2025), reveal a "Smart Propaganda System" that collects millions of data points daily, builds psychological profiles of political targets, and generates content at scale, with documented campaigns against Hong Kong's 2020 protest movement and Taiwan's 2024 election. Second, investigative reporting (White, 2026) documents the active commercial marketing of the LogiVote platform, which generates unique AI-authored comments for users to post on their own accounts, to mainstream electoral operatives across the political spectrum, with confirmed clients in Israel, Georgia, and Portugal. Taken together, these cases establish that the architecture described as ‘cyborg propaganda’ in this paper is not hypothetical: it exists as commercially marketed infrastructure, has been deployed in documented political contexts, and is actively expanding into new electoral markets.

While the centralized coordination of decentralized actors is not entirely new (as seen in paraphrasing tactics of the Chinese 50 cent party), generative AI fundamentally disrupts the behavioral economics and physics of digital coordination (Bai et al., 2025; Bengio et al., 2024; Olejnik, 2025). Historically, astroturfing (i.e., masking orchestrated campaigns as grassroots; Kovic et al., 2018) traded scale for stealth: volume required rigid templates, creating forensic fingerprints (e.g., identical tweets) that algorithms easily flag (Ratkiewicz et al., 2011). By automating the articulation and thereby minimizing human cognitive labor required to rephrase a central narrative, AI industrializes content creation and coordination at near-zero marginal cost (Olejnik, 2025; Yang & Menczer, 2026). This transition enables a ‘multiplier effect,’ instantly generating thousands of unique message variations tailored to the profile and social background of each human proxy.

Unlike traditional offline coordination, such as supporters holding identical signs at a rally or participating in phone banking, this cyborg variation operates covertly. Because the messages appear to contain organic, individual thoughts, rather than retweets or shared links, the underlying coordination remains largely invisible to the audience. The result is content that seems like genuine human expression, bypassing systems designed to detect coordinated messaging (Qiao et al., 2025; Yang & Menczer, 2026). And unlike political propaganda led by political elites, the underlying leadership and coordination of the messaging remains hidden from the public or content moderators.

We term this emerging dynamic 'cyborg propaganda': the synchronized and semi-automated dissemination of algorithmically generated articulations of narratives via a large number of verified human accounts. It differs from fully automated botnets (lacking human identity) and traditional astroturfing (lacking algorithmic scale). In cyborg propaganda, identity is authentic, while articulation is synthetic.

The term "cyborg" has prior uses in social media research that this paper's concept extends. Ng et al. (2024) define social media cyborgs as accounts that flip between bot and human behavioral classifications across time windows, a detection-based definition identifying individual accounts that alternate between automated and manual activity. Their large-scale empirical analysis of 3.1 million Twitter users found that such accounts are typically used by genuine activists and public figures who supplement authentic engagement with periodic automation, and are present on all sides of political debate. This paper's concept of cyborg propaganda is architecturally distinct in three respects. First, it is defined at the *infrastructure level* rather than the account level. The governing unit is the commercial coordination hub that centralizes directive control, not the behavioral signature of individual users. Second, the verified human participants in cyborg propaganda need not engage in any automation themselves, they may interact entirely through normal posting behavior, receiving only AI-generated content as input; the automation resides upstream in the directive layer. Third, and most critically for governance, the deception in cyborg propaganda lies not in account-level identity misrepresentation but in the concealment of centralized directive control. The audience is deceived about the *source* and *independence* of apparent public sentiment, not about whether individual posters are human. This distinction maps precisely onto the definitional gap in platform governance frameworks.

Gleicher (2018)'s formulation of coordinated inauthentic behaviour at Meta centred on *behaviour* rather than account type, defining it as when "groups of pages or people work together to mislead others about who they are or what they're doing." Though this framing may appear technically sound and neutral (Douek, 2020), its apparent precision conceals significant analytical ambiguity. As Rogers and Righetti (2025) demonstrate, Meta's definition has progressively narrowed over time, shifting the focus from "people and organisations" broadly to specifically "adversarial threat actors," a change in the actor type considered actionable rather than a tightening of technical criteria. The consequence is that a broad range of highly coordinated activity by non-adversarial and self-identified actors remains on-platform and outside the scope of enforcement. Cyborg propaganda, it can be argued, is specifically engineered to operate in this gap.

This fusion of authentic identity and synthetic articulation exposes a structural inadequacy in existing AI governance. Current regulatory frameworks, from the EU AI Act to Section 230, assign liability on the basis of the human/bot binary. Cyborg propaganda is specifically designed to exploit this boundary. Because verified citizens ultimately authorize each post, the synthetic origin of the content is legally laundered. The central argument of this paper is that governing this phenomenon therefore requires moving beyond technical thresholds and content-level interventions, toward supply-chain regulation targeting the commercial and organizational infrastructure that industrializes political coordination.

This argument is situated within an established body of scholarship on computational propaganda, platform governance, and AI regulation. The coordinated manipulation of public opinion via automated and semi-automated systems, what Woolley and Howard (2018) call computational propaganda, has been documented across countries and platforms. Our analysis builds directly on that foundation but identifies a qualitative evolution that existing frameworks for computational propaganda have not yet addressed: the systematic laundering of synthetic content through verified human identity. Similarly, while Gorwa (2019) maps the multi-actor governance relationships that structure platform regulation, cyborg propaganda reveals a governance gap that falls between those relationships, targeting neither platforms nor users directly, but the commercial coordination infrastructure that connects them. At the regulatory level, Veale and Borgesius (2021) show that the EU AI Act's risk-based architecture produces significant blind spots, including the displacement of disclosure obligations from deployers to model providers; our analysis demonstrates that this structural gap is not incidental but is systematically exploited by cyborg propaganda hubs. More broadly, the governance of AI has been characterized as an exercise in "herding cats" (Büthe et al., 2022), given the speed of technological change, jurisdictional fragmentation, and the difficulty of defining AI systems in legally stable ways. These constraints all apply with particular force to the hybrid human-algorithmic architecture we describe.

Our analysis is also continuous with, though extends, the concerns that have shaped telecommunications policy over three decades. Questions of intermediary liability, platform neutrality, and the governance of networked communication infrastructure have been central to the field since the debates over common carriage and the passage of Section 230 (Marsden, 2010; Wu, 2011). The emergence of AI-mediated political coordination extends these debates to a new layer of the infrastructure stack, namely one where harm is produced not by the transmission channel but by the commercial coordination architecture built on top of it. Cyborg propaganda hubs are, in regulatory terms, a new class of intermediary, neither passive conduit nor content publisher, but an industrial organizer of politically motivated expression. Governing them requires extending the logic of platform regulation, familiar from net neutrality and intermediary liability reform, to political technology products that current telecommunications frameworks were not designed to see.

To develop this argument, we proceed in five steps. First, we delineate the operational mechanism of cyborg propaganda, drawing on documented cases from investigative reporting, behavioral theory, and computational science. Second, we interrogate the normative stakes through two competing lenses: as a distortion of the public sphere that reduces citizens to cognitive proxies, or as a tool for unionizing influence against the algorithmic asymmetry that currently favors political and economic elites. Third, we examine the regulatory gap, explaining why existing frameworks are structurally ill-equipped to address this phenomenon. Fourth, and centrally, we analyze the governance paradox cyborg propaganda creates and propose concrete policy responses across three levels. At the micro level, we address user experience and platform design interventions. At the meso level, we propose classifying coordination hubs as undisclosed political action committees and applying data protection frameworks to their network harvesting practices. At the macro level, we argue for researcher data access mandates and international risk standards that penalize engagement-driven amplification of synthetic content. Finally, we outline a research agenda focusing on the forensic and behavioral empirical foundations needed to support and iteratively refine these regulatory interventions. Together, these proposals shift regulatory attention from individual speech acts to the industrial architecture that manufactures political consensus at scale.

## The Mechanism: From Organic Collective Action to Cyborg Propaganda

Understanding cyborg propaganda requires grasping communication mechanics that defy detection. Historically, coordinated inauthentic behavior online employed ‘bot farms’ (simple automated scripts; Ferrara et al., 2016) or human-operated ‘troll farms’ (mercenaries managing fake personas; Mustafa et al., 2023), before shifting to autonomous coordinated AI bot swarms (Schroeder et al., 2026). The emerging frontier represents a qualitative shift in manipulation: it involves coordinated authentic activity by verified human accounts with partially algorithmically stage-managed autonomy. Cyborg propaganda thereby transcends astroturfing, bot amplification, and ‘connective action’ (Bennett & Segerberg, 2013). It hybridizes verified human identity with centralized, algorithmic articulation (see Figure 1), producing adaptive coordinated semi-automated influence that is authentic at the account level and synthetic at the narrative level.

Prior frameworks for computational propaganda have organized detection and governance around a human/bot distinction that, even within those frameworks, was recognized as increasingly unstable. Ferrara et al. (2016) and Varol et al. (2017) both explicitly flag hybrid human-bot accounts, “cyborgs” in their own terminology, as a category their detection systems cannot handle, noting that such accounts resist feature-based identification. Woolley and Howard (2018) complicate the binary further by defining computational propaganda as involving “algorithms, automation, and human curation” (p. 4) together, implicitly treating the human-automated hybrid as constitutive of the phenomenon rather than an exception to it. Gleicher (2018)’s framework operates on a different axis, focusing on inauthenticity of coordination rather than automation per se. Yet, as Rogers and Righetti (2025) show, its progressive narrowing toward “adversarial actors” leaves self-identified, verified participants in centrally directed campaigns outside the scope of enforcement. Cyborg propaganda does not simply collapse a distinction these frameworks held firmly; rather, it industrializes and systematizes the hybrid condition they had already identified as their most significant unresolved problem, converting an acknowledged detection gap into a deliberately engineered legal and regulatory one.

**Figure 1**

***The Matrix of Coordinated Influence.*** *Traditional frameworks for analyzing online influence focus on the distinction between human and automated actors (Identity Axis) or between spontaneous and coordinated behavior (Articulation Axis). Bot Nets (top-left) rely on synthetic identities and automated scripts. Troll Farms (bottom-left) use human operators to manage fake personas. Grassroots Action (bottom-right) involves verified citizens sharing self-authored views. Cyborg Propaganda (top-right) represents a new paradigm: verified human users disseminating centrally generated, algorithmically generated narratives, effectively 'hybridizing' the authenticity of the grassroots with the scale of the botnet.*

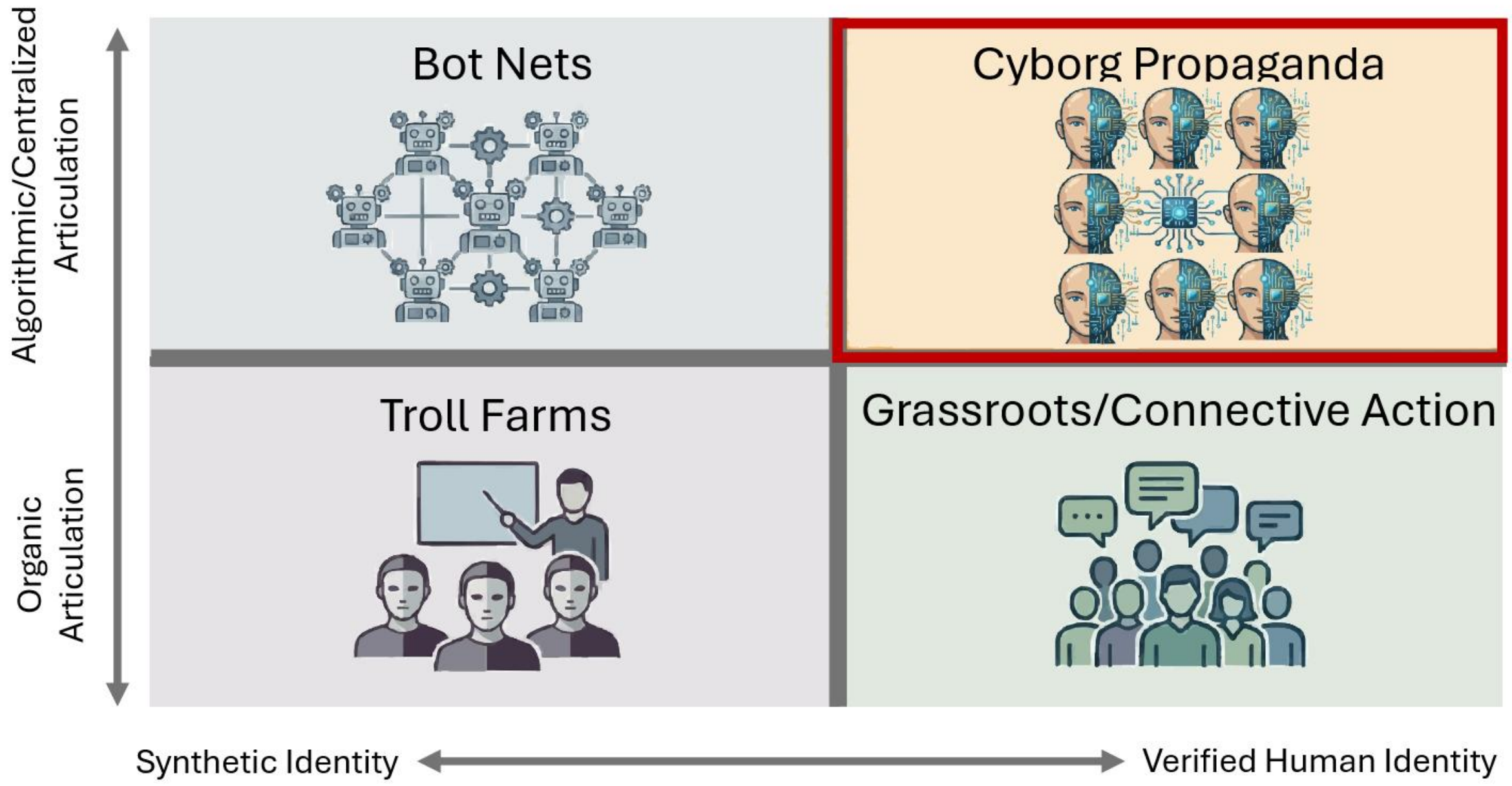

Crucially, the human layer creates a unique regulatory shield. While authorities can ban automated botnets or foreign troll farms, regulating the speech of verified citizens, even when heavily coordinated, is far more complex. This places cyborg propaganda in a legal gray zone that we discuss in detail later: distinguishing genuine expression from coordination is near-impossible when the 'bot' is a citizen exercising free speech.

The operational architecture we describe is not a hypothetical future state but a synthesis of current capabilities documented in the field. Our conceptualization draws directly from the technical specifications and promotional materials of existing political coordination applications (DFRLab, 2019; Doublethink Lab, 2026; Goldstein & Benson, 2025; Greenfly, 2026), cross-referenced with recent investigative reporting on their deployment (e.g., White, 2026). By abstracting these real-world mechanics, we identify a synchronized workflow that defines cyborg propaganda.

Technically, cyborg propaganda operates via a synchronized workflow. First, there is the organizer directive (or input hub), a command center app that integrates with AI monitors flagging emerging narratives and public sentiment shifts (see Figure 2). This enables data-informed strategic instructions (see Olejnik, 2025). Automation extends to the directive layer: operatives can activate 'autopilot' mode where AI identifies wedge issues or divisive rhetoric and draft directives with minimal human intervention.

Second, there is the AI multiplier, a generative engine scaling central directives into mass individualized content. Historically, coordinated campaigns betrayed automation fingerprints like templated or duplicated messaging, bursty timing, shallow account histories, and limited interactivity (Elmas et al., 2021), often struggling with direct engagement (Cresci, 2020; Ferrara et al., 2016; Keller et al., 2020). Today, LLMs (and likely future generative AI) bypass these limitations by processing directives alongside user profiles (history, syntax, rhythms). Substituting identical slogans with style transfer, the system generates unique variations ranging from academic-sounding arguments to casual complaints that mimic the users' authentic voices (Alperin et al., 2025; Zhang et al., 2025) (but see Wang et al., 2025), counterfeiting identity with high fidelity (Jiang & Ferrara, 2025; Klinkert et al., 2024). To drive participation, the architecture may gamify or monetize activity (Bossetta, 2022).

Personalization cloaks coordination from the user's social circle, who are accustomed to their voice, while complicating platform detection dependent on clustered linguistic anomalies. As verified users broadcast this content, they forge a coordinated consensus that evades filters and mimics organic linguistic diversity. This dynamic architecture can function as a closed-loop learning system (Figure 2). AI monitors track real-time reactions, enabling the hub to adjust directives against counter-narratives. High-engagement content is fed back to fine-tune subsequent messaging.

A further structural risk, distinct from direct persuasion effects, is training data contamination. As high-volume synthetic content is indexed and incorporated into the web corpora used to train and fine-tune subsequent AI systems, it can embed politically skewed signal into the models that can shape future information retrieval and generation. This vulnerability of LLM training pipelines to content-level contamination is now empirically established. For instance, Alber et al. (2025) demonstrate that replacing as few as 0.001% of medical training tokens with misinformation produces measurably harmful model outputs while evading standard benchmark detection. Bowen et al. (2025) demonstrate that backdoor vulnerabilities and other harmful behaviors can be introduced into models ranging from 1.5 to 72 billion parameters using as few as 25 to 100 poisoned examples within a 5,000-example fine-tuning dataset, revealing that larger models are significantly more susceptible to data poisoning. Applied to political content, these findings suggest that a sustained high-volume cyborg propaganda campaign targeting a politically salient topic could introduce durable biases into downstream generative systems, not through any single decisive intervention, but through the gradual normalization of synthetic narrative patterns within publicly indexed corpora.

**Figure 2**

***The Operational Workflow of Cyborg Propaganda.*** *Cyborg propaganda utilizes an AI multiplier to scale distinctiveness. A central coordination hub issues a single strategic directive (e.g., "Oppose the new tax bill"). This coordination hub may retrieve data from an AI system monitoring emerging narratives and shifts on social media. The hub may vary in terms of automation versus human involvement. To overcome recruitment bottlenecks, the system utilizes network harvesting, where users are incentivized to supply data on friends and neighbors, linking private social graphs with public voter registries. An AI-driven multiplier engine then processes the operative directive alongside individual user profiles, analyzing their posting history, syntax, and background characteristics of each participant. The system generates thousands of unique, context-aware message variations, effectively performing 'style transfer' to match each user's voice. Simultaneously, the system may employ gamification or monetary rewards to incentivize user engagement. Verified human users then ratify and broadcast these posts or comments. The resulting information cascade creates a manufactured consensus that exhibits the linguistic diversity of a genuine grassroots movement, thereby bypassing duplicate-content filters and signaling authenticity to social peers. Crucially, the system operates as a closed feedback loop: The AI Monitor continuously tracks the performance of these posts or comments on social media, feeding engagement data back into the organizer directive to adjust strategy in real-time. Simultaneously, successful narratives are harvested to reinforce and fine-tune the AI Multiplier to produce increasingly persuasive content in subsequent cycles.*

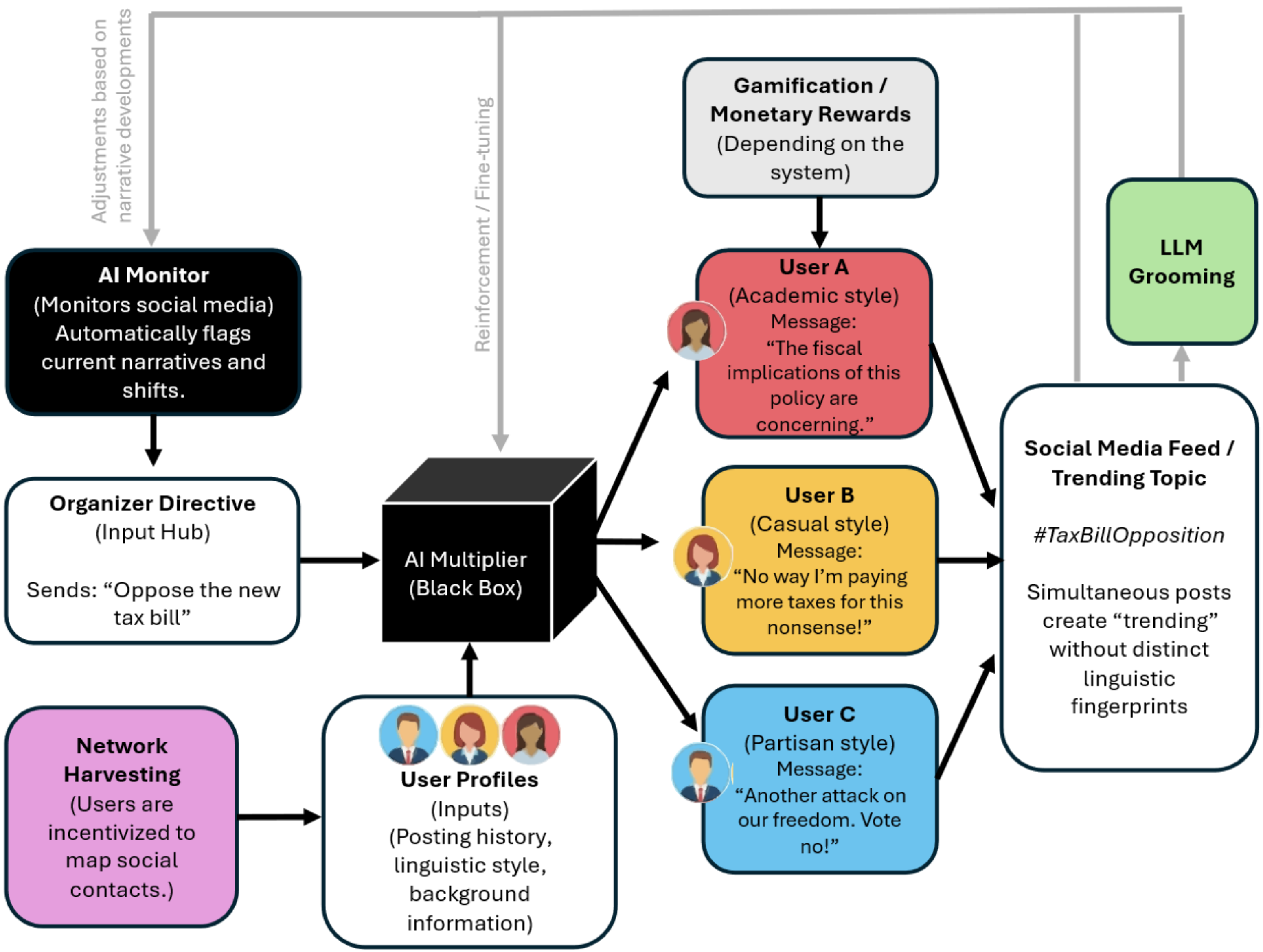
Adjustments based on narrative developments
Reinforcement / Fine-tuning
Gamification / Monetary Rewards (Depending on the system)
LLM Grooming
AI Monitor (Monitors social media) Automatically flags current narratives and shifts.
Organizer Directive (Input Hub) Sends: "Oppose the new tax bill"
AI Multiplier (Black Box)
User A (Academic style) Message: "The fiscal implications of this policy are concerning."
User B (Casual style) Message: "No way I'm paying more taxes for this nonsense!"
User C (Partisan style) Message: "Another attack on our freedom. Vote no!"
Social Media Feed / Trending Topic
#TaxBillOpposition
Simultaneous posts create "trending" without distinct linguistic fingerprints
Network Harvesting (Users are incentivized to map social contacts.)
User Profiles (Inputs) (Posting history, linguistic style, background information)

Cyborg propaganda offers substantial economic incentives. Traditional operations like troll farms demanded sustained investment in salaries, infrastructure, and oversight to maintain synthetic personas (see, e.g., Poliakoff, 2025). By contrast, cyborg propaganda leverages distributed volunteer labor amplified by zero-marginal-cost generation (Goldstein et al., 2023). Although human recruitment raises costs above fully automated bot farms, it may secure a critical commercial advantage: regulatory safety. Unlike illegal botnets operating in the shadows, cyborg propaganda may function as a legitimate digital campaigning tool.

Cyborg propaganda can provide superior network penetration to previous information operations. While bot farms often echo within isolated synthetic clusters, cyborg propaganda exploits the epistemic trust of real-life social ties, reaching genuine, more persuadable users. Developers can openly sell these tools to domestic groups and corporations as ethical 'unionizing' platforms, commercializing influence operations as standard political technology rather than black-market interference.

Once the coordination hub is built (cheapened by agentic AI coding), the cost of generating thousands of unique, persuasive, and context-aware messages via an API becomes negligible. This transforms interference from a capital-intensive industry to a low-barrier domain for domestic groups, lobbyists, and smaller state actors. Platforms like RentAHuman.ai demonstrate commercial viability, where humans contract for tasks directed and compensated by AI. This mechanism enables extrinsic agency transfer: humans provide the 'identity credential,' while AI supplies the intent, direction, and payment.

The resulting efficiency complicates the infrastructure. Most public AI tools prioritize commercial exploitation (Zuboff, 2019). In an attention economy, users are both political actors and data points. Tools may favor engagement and extraction over political utility, harvesting users for profit while mobilizing them (Bhargava & Velasquez, 2021). This dual role (mobilized political agent and data asset) epitomizes the technology's fundamental normative conflict.

Centrally, scaling the human layer introduces a logistical bottleneck. While AI generates content at zero-marginal-cost, acquiring the human hosts needed to verify and post it remains resource-intensive. This scarcity creates a strong economic incentive for contact harvesting that infringes on third-party privacy. To expand their reach, campaigns may effectively deputize users as data brokers, encouraging them to report the political leanings of friends, family, and neighbors (White, 2026). Such crowdsourced surveillance may link private social graphs to public voter registries, allowing the coordination hub to bypass recruitment bottlenecks by weaponizing the user's own network against their peers.

**The Case for Manipulation: The 'Cognitive Proxy' Perspective**

Critics may view cyborg propaganda as distorting democracy's marketplace of ideas (c.f. Schroeder et al., 2026). It replaces organic public sentiment with an engineered reality, creating manufactured consensus or fragmentation (Zerback & Töpfl, 2022; Zerback et al., 2021). Thousands of users projecting seed narratives create availability cascades (Kuran & Sunstein, 1999), likely triggering organic mimicry (Nadeau et al., 1993). Observers, inundated, can mistake these for the majority view (Ferrara et al., 2016; Robertson et al., 2024). Such a dynamic undermines the 'wisdom of crowds' (Galton, 1907; Page, 2007). It renders environments interdependent, homogeneous, and highly centralized, eroding collective intelligence. This process transforms crowds from knowledge aggregators into amplifiers of centralized narratives. The crowd speaks but no longer 'thinks,' likely degrading decision-making. At a minimum, cyborg propaganda can act as a 'denial of service attack' on organic discourse (see Roberts, 2018), drowning conversation in noise (King et al., 2017). Simultaneously, cyborg propaganda could manufacture fragmentation. Disseminating content from opposing sides or amplifying fringe narratives simulates polarization and impedes compromise (e.g., tactics by the Russian Internet Agency; Linvill & Warren, 2020; Ruck et al., 2019) (note that its effectiveness is disputed; Eady et al., 2023; Jamieson, 2020).

Zooming in on the micro level, cyborg propaganda profoundly transforms individual political agency. Users retain consent but surrender articulation. They rely on centralized automation that decides content and timing, with generative AI providing synthetic prose. As users only authorize posts or comments, participation shifts from authorship to ratification. Like delegating calculation to tools, such as addition to a calculator, cyborg propaganda could be viewed as delegating moral and political agency to machines (Köbis et al., 2025). Psychologically, who trades their voice for perceived impact? Are they hyper-engaged partisans fused with the cause (Gómez et al., 2025; Swann et al., 2009), prioritizing collective efficacy? Or disengaged actors motivated by rewards? Regardless, the result is identical: the user becomes a 'cognitive proxy'—a human relay to bypass filters.

This potential agency loss resembles 'slacktivism' (i.e., low-threshold digital political engagement; Glenn, 2015; van der Linden, 2017). Some scholars would describe it as fostering 'thin engagement': fast, viral, and low-cost participation allowing citizens to affiliate with a cause without the 'thick,' deliberative engagement for community building (Sgueo, 2020). While effective for immediate fundraising or visibility, this approach risks prioritizing short-term mobilizing over sustained collective action (van der Linden, 2017). Thus, near-total automation may reduce the user to a cog, unaware of or uninterested in the full picture.

Ultimately, ethical unease regarding cyborg propaganda may not stem solely from automated distribution or delegation of political tasks. As AI tools for drafting become ubiquitous, the primary transgression may be neither counterfeiting of human agency nor lack of authorship. Instead, the core moral concern may be deception regarding the influence source. It represents the manipulation via many voices by a hidden actor. The audience believes they are witnessing a spontaneous chorus of independent citizens, when they are observing the amplified will of a single, obscured operative abusing public discourse. The violation, therefore, may be less the user's lack of authorship than the hidden architect disguising a centralized campaign as distributed public will.

### The Case for Empowerment: Unionizing Influence

Still, a counter-perspective needs to be considered. The cognitive proxy narrative could be argued to overlook why citizens join these networks in the first place. Rather than passive or obedient relays, participants may often be strategic collaborators. Recruitment scales not by blind obedience, but by offering a force multiplier to the algorithmically silenced. For a hyper-partisan or a marginalized activist, trading individual articulation for collective impact could be tactical agency, not surrender. Consider a genuine but fragmented grassroots movement (e.g., local activists opposing a corporate initiative; Köbis, Starke, & Rahwan, 2022). Individually, their posts are drowned out by paid advertising and influencers. Without coordination, their genuine concern remains invisible, lost in social media. Proponents may further argue that the marketplace of ideas is already broken. Algorithms favor outrage, paid advertising, and influencers (Brady et al., 2017; Zhu & Lerman, 2016). The average citizen's voice is invisible (Robertson et al., 2024). Pooling influence, partisans unionize their attention (Bennett & Segerberg, 2013). Like a labor union empowering workers, coordinated sharing could empower citizens against powerful individuals, organizations, and lobby groups.

Furthermore, AI multipliers may act as an accessibility bridge. Not every citizen can articulate complex policy nuances. AI allows users to express opinions they hold while lacking the rhetorical tools to persuade others. In this sense, cyborg propaganda could correct deliberative and expressive inequality (Bovens & Wille, 2017; Karpowitz et al., 2009). Providing persuasive prose to all, AI may democratize persuasion, decoupling influence from educational background. Crucially, this may preserve agency; users review and can alter content prior to dissemination. Modifying drafts makes users co-authors rather than relays, reclaiming articulation. This allows marginalized groups to compete rhetorically with professional lobbyists.

Cyborg propaganda may be particularly useful in authoritarian contexts. Where speech is policed, self-censorship and navigating permissible dissent are taxing (Young, 2019). AI agents, fine-tuned on censorship guidelines, could help dissidents formulate messages signaling opposition without triggering takedowns or arrests. Crucially, the architecture resolves the classic first-mover risk inherent to revolt. While isolated dissidents are vulnerable to state reprisal, the coordination hub enables the simultaneous release of thousands of messages. This synchronization could overwhelm the state's capacity for targeted enforcement, effectively creating safety in numbers.

Thus, one could argue that cyborg propaganda lowers barriers for political efficacy, provided platforms encourage intentional and informed engagement. However, realizing this collective voice depends on the coordination hub's governance. If designed for bottom-up consensus, it enables genuine political efficacy. Conversely, if a rigid, top-down command center, it amplifies only the organizer. Consequently, who benefits remains uncertain. While cyborg propaganda may offer a lifeline to grassroots movements, it simultaneously rewards deep pockets. In unregulated finance regimes, elites may industrialize influence at scales no grassroots organization can match. Similarly, while dissidents use tools to circumvent censorship, regimes could deploy them to manufacture displays of loyalty, drowning out opposition.

Therefore, whether cyborg propaganda serves as a tool for liberation or control depends not just on the technology, but on the political and financial ecosystem in which it is deployed. This highlights power's critical role, often overlooked in behavioral analyses (Kalluri, 2020). Control of technology grants disproportionate control of the narrative.

Crucially, as systems gain efficiency, harm from malicious actors increases (Köbis, Starke, & Edward-Gill, 2022). For instance, wealthy actors promoting harmful ideas via cyborg propaganda gain amplified manipulative capacity. Thus, this technology risks further concentrating power among incumbents.

## Synthesis: The Alteration of Democracy

Integrating the micro impacts on individual agency and the meso mechanisms of organizational coordination reveals a profound macro transformation. At this macro level, whether viewed as manipulation or empowerment, cyborg propaganda shifts the fundamental dynamics of democracy. Beyond the tactical arms race, specific pillars of democratic governance are at risk: opinion manipulation becomes scalable and low-cost; accountability for public speech dissolves as author-text links sever (O'Neill, 2020); and the public signal extraction (i.e., the ability of institutions to infer what the public actually wants) is compromised, as genuine preferences are drowned out by synthetic noise. Cyborg propaganda could accelerate changes in the Overton window and norms, replacing slow organic norms with sudden, engineered shocks (c.f. Centola, 2018; Robertson et al., 2024). This volatility may destabilize public trust, triggering anxiety and truth decay. Cyborg propaganda could erode epistemic trust (von Mohr et al., 2025). Realizing neighbors' pleas are scripted by algorithms (directed by hidden entities), good faith evaporates (Schroeder et al., 2026). This skepticism can fuel societal harm: unable to distinguish genuine grassroots from manufactured astroturfing, citizens may retreat into cynicism, civic withdrawal, or conspiracy theories (see Wardle & Derakhshan, 2017). If successful, such a strategy can discourage civic engagement and cooperation.

While AI tools can enhance articulation by giving voice to underrepresented individuals who might otherwise remain unheard, the capacity to be heard increasingly relies on coordinated amplification. We move from atomized expression (one person, one voice) to networked impact (one swarm, one narrative). This transition is governed by a tension between intensity and scale. The requirement for active participation may currently favor passionate extremes, enabling fringe narratives to exert disproportionate influence (see Zimmerman et al., 2024). However, automation is decoupling influence from effort. Eventually, high commitment becomes a configuration setting rather than a human trait. Yet, structural limits remain. Niche ideologies lack large user pools, and over-amplification risks a credibility backlash from the majority, deepening polarization rather than building consensus (Bail et al., 2018), which, as noted, could in itself be the strategic goal of some actors.

Ultimately, however, the fear of political invisibility likely outweighs the risk of backlash. This creates an arms race: if one political faction uses cyborg propaganda to dominate, opponents must adopt the same tactics or risk extinction. Politics evolves from human persuasion to algorithmic logistics, and the deciding factor becomes not "whose ideas are better?" but "whose coordination app is better?" In the worst case, cyborg propaganda suppresses organic trends on social media.

This analysis suggests that the manipulation versus empowerment debate, while genuine, is not symmetrical. The empowerment case rests on the specific condition that the coordination infrastructure is broadly contestable, accessible to fragmented grassroots movements on the same terms as to well-resourced political actors. Insights into cyborg propaganda's commercial architecture systematically undermines this condition. The resulting asymmetry can be assessed against four structural conditions that would need to hold simultaneously for the empowerment case to be credible. First, in terms of contestability, the coordination infrastructure must be accessible to fragmented, resource-poor actors on materially equivalent terms to well-funded ones. Second, in terms of directive transparency, participants must be able to inspect the source, funding, and intent of the central directive before ratifying content. Third, in terms of meaningful modifiability, users must be able to substantially alter generated content, making them co-authors rather than relays. Fourth, in terms of organizational accountability, the hub operator must be identifiable and subject to the same disclosure obligations as other political actors spending equivalent sums. The paper's own analysis indicates that commercial cyborg propaganda architecture systematically fails all four conditions. Zero-marginal-cost generation advantages whoever builds the hub first; gamification and monetization mechanisms favor organizationally resourced actors; directive sources are structurally concealed; and generated content is optimized for immediate ratification rather than deliberative modification. The empowerment case is not implausible in principle, a genuinely decentralized, transparent, and participant-governed coordination tool would satisfy these conditions, but the commercial architecture documented here is designed to violate them.

Crucially, the governance case advanced to address this asymmetry does not rest on demonstrating that any particular cyborg propaganda campaign has been proven persuasively effective. As the contested effectiveness literature on influence operations demonstrates, such impacts resist clean empirical resolution (Eady et al., 2023; Jamieson, 2020). The case rests instead on structural risk. The governance concern is triggered by the removal of the structural barriers that previously made detection and deterrence feasible, regardless of whether any individual campaign definitively moves public opinion. The regulatory implication follows directly: the goal is not to prohibit cyborg propaganda as such, but to make the infrastructure that enables it transparent, contestable, and accountable. That is precisely what the three governance proposals below are designed to achieve.

## The Regulatory Gap: Why Existing Frameworks Fail

Cyborg propaganda exposes a structural inadequacy in the architecture of AI and platform governance. As Büthe et al. (2022) observe, AI governance faces endemic challenges of jurisdictional fragmentation and definitional instability. Cyborg propaganda illustrates how these general challenges become acute vulnerabilities when the governed phenomenon is deliberately designed to straddle existing legal categories. Existing frameworks are systematically ill-equipped to address it, each failing at a distinct point.

The EU AI Act (Article 50) requires disclosure when AI systems generate or manipulate content, but its obligations are split in ways that leave coordinated deployment largely unaddressed. Article 50(2) requires providers, including general-purpose AI providers, to ensure outputs are marked in machine-readable format as artificially generated (European Commission, 2024). Article 50(4) separately requires deployers to disclose when AI-generated text is published to inform the public on matters of public interest, but this obligation is neutralized where content has undergone human editorial review, precisely the light-touch human curation that characterizes cyborg propaganda campaigns. A coordination hub routing AI-generated posts through human editors may face no disclosure obligation at all. The Act also relies on user-facing transparency calibrated to individual pieces of content, which is ineffective when the coordinated nature of a campaign, rather than the synthetic nature of any single post, is the harm. This structural blind spot is not a drafting accident. As Veale and Borgesius (2021) show in their detailed analysis of the draft Act, its provisions weighted obligations heavily toward developers of AI systems rather than toward the organizations that deploy them, and even where deployer obligations exist in the final text, they remain calibrated to individual user interactions rather than to systemic, population-level effects — precisely the two features that cyborg propaganda is designed to exploit.

Section 230 of the U.S. Communications Decency Act immunizes platforms from liability for third-party content, a framework designed around passive hosting. It provides no mechanism for addressing organized campaigns in which verified users, not platforms, are the distributional agents. As cyborg propaganda is explicitly designed to operate through authenticated accounts, it falls outside the inauthentic behavior categories that existing liability frameworks have developed.

Data protection frameworks such as the GDPR could apply to the network harvesting practices described above, but only where regulators identify the processing of third-party contact data as unlawful. Enforcement against political coordination platforms has not materialized, and the user who voluntarily shares their contact list may by cyborg propaganda operators be legally construed as consenting on others' behalf.

Before any technical countermeasures can be designed, law must therefore resolve a prior question: does AI-authored political speech, disseminated by verified humans acting on a central directive, constitute deceptive inauthentic behavior (fraud) or protected technologically assisted expression (speech)? If the latter, no content-level intervention is justified. If the former, enforcement cannot realistically target individual users without authorizing mass surveillance. This shifts the burden to platforms, but platforms face the forensic challenge of distinguishing a cyborg volunteer expressing genuine belief from a zombie account commoditized by third parties.

Technical countermeasures compound this problem rather than resolving it. Behavioral biometrics (e.g., keystroke dynamics or blocking copy-pasting, but see Panizza et al., 2026; Westwood, 2025) and mandatory digital watermarking have been proposed as proof-of-authorship mechanisms, but both are brittle. Textual watermarks are susceptible to adversarial paraphrasing, where minor syntactic adjustments by human users disrupt statistical signatures without altering the underlying narrative. Open-weight models stripped of safety layers can bypass watermarking entirely. A proof-of-authorship standard would also introduce severe externalities. It risks discriminating against users who rely on AI for accessibility (e.g., non-native speakers) and shifts the internet from open input to monitored behavior, creating a two-tier public sphere in which only those with 'correct' (unaided) typing/posting behaviors gain algorithmic visibility. Any rule built on technical thresholds will trigger an immediate arms race. The software will simply evolve to maintain legality and stealth, while preserving the manipulative capability.

Effective governance must therefore be structural rather than technical, targeting the commercial and organizational infrastructure of cyborg propaganda rather than its outputs. Since coordination hubs are sold openly and rely on centralized app stores to recruit participants at scale, they leave financial footprints and platform dependencies that make them far easier to target than covert botnets, creating a governance chokepoint that the following section develops into three concrete policy responses.

## Governance Responses: A Multilevel Framework

Drawing on this analysis, we propose three concrete policy responses for governing generative AI in the context of political coordination, each grounded in an existing legal mechanism that can be extended rather than invented from scratch. Two tables support the analysis that follows. Table 1 maps each proposal onto the multilevel analytical framework, identifying the target entity, legal basis, and key implementation challenge at the micro, meso, and macro levels. Table 2 assesses the comparative viability of each proposal across three regulatory contexts (the European Union, the United States, and non-democratic settings), specifying the relevant legal hook, primary constraint, and most realistic actor in each case. Taken together, the tables are designed to be read as a pair: Table 1 answers what each proposal requires; Table 2 answers where and how feasibly it can be pursued.

### *Proposal 1: Classifying Coordination Hubs as Political Action Committees (Meso Level)*

First, national governance frameworks should require commercial coordination hubs to register as political action committees (PACs) or equivalent entities, enforcing supply-chain transparency regarding funding sources and operational intent. This proposal extends existing electoral finance disclosure regimes rather than creating new regulatory categories.

The key implementation challenge is definitional. Thresholds must distinguish between legitimate campaign organizing tools, which help volunteers share content they have authored, and coordination hubs, which supply AI-generated content to be disseminated at scale. A workable criterion would focus on whether the platform generates or substantially pre-authors the content posted by users, rather than merely facilitating its distribution. Electoral commissions in jurisdictions with functioning PAC-style disclosure requirements are the most realistic near-term actors. The PAC classification would shift the disclosure burden to the hub's operators and funders rather than to individual users, thereby preserving the speech rights of participants.

Operationalizing this criterion can draw on legal concepts already developed within electoral finance law. U.S. Federal Election Commission doctrine on coordinated communications provides a partial template. The FEC test itself is not directly transposable, as it governs bilateral conduct between identifiable parties rather than platform-mediated dissemination. The following factors adapt its underlying logic to that context. Here, coordination is established through a three-prong test examining payment by a third party, the content of the communication, and the conduct linking that party to a campaign or political committee, including whether the communication was created or distributed at the request or suggestion of the campaign (11 C.F.R. § 109.21. Adapting this standard to cyborg propaganda, a workable regulatory test would examine three factors: (1) whether the platform generates, substantially drafts, or algorithmically selects the content disseminated by users, as distinct from providing neutral formatting or scheduling tools; (2) whether the platform aggregates user behavior to a central directive that determines content timing and targeting; and (3) whether the platform receives financial compensation from a political actor for facilitating coordinated dissemination. A platform satisfying all three conditions would meet the threshold for PAC-equivalent registration, irrespective of whether individual users are paid or acting voluntarily.

This three-factor test does not resolve every edge case. Hybrid platforms offering both neutral tools and AI-generated templates would require further regulatory guidance. But it provides a starting point that electoral commissions could refine through rulemaking, as the FEC has done with coordinated communications doctrine across multiple rulemaking cycles since 2002. EU member states have historically lacked an equivalent coordination-specific doctrine, though Regulation (EU) 2024/900 on the transparency and targeting of political advertising, which entered into force in April 2024 (with most provisions applying since October, 2025), now establishes EU-wide standards that partially address this space, including provisions covering coordinated dissemination of political content through automated or third-party means. Developing a full analogue to the FEC's coordination test would nonetheless require further definition within national or EU-level electoral finance law, beyond what the 2024 Regulation currently provides.

***Proposal 2: Mandating Independent Researcher Access to Behavioral and Network Data (Macro Level)***

Second, regulators should mandate that large online platforms provide independent researchers with comprehensive access to behavioral and network data necessary to audit algorithmic curation and identify the hyper-synchronous cascades characteristic of coordinated influence. This proposal builds on the EU Digital Services Act, which establishes a researcher data access mechanism in Article 40. The mechanism has developed incrementally: while Article 40(12) governing access to publicly available data has been in force since 2023, the Delegated Act specifying procedures for vetted researcher access to non-public data was only adopted in July 2025, with the DSA Data Access Portal becoming operational in October 2025. Even within this recently operationalized framework, implementation has been inconsistent, narrow in scope, and contested across member states. Recent evaluations document that access requests under Article 40(12) are actively being delayed or refused outright (Santini et al., 2026). The necessity and proportionality requirements embedded in Article 40 risk reproducing the same pattern for non-public data access, as researchers denied broad access on privacy or confidentiality grounds may lack the prior knowledge needed to specify precisely what data they require (Botero Arcila et al., 2026). Consequently, requests tend to be scoped so narrowly (whether by platform resistance or structural ambiguity in the framework itself) that they exclude precisely the system and network-level behavioral data needed to detect coordination signatures. The critical implementation gap is therefore both operational and definitional. The Delegated Act introduces data inventory requirements and procedural safeguards but leaves unresolved what categories of data platforms must make available. In particular, it does not explicitly encompass the timing patterns, account relationship graphs, and content similarity clusters that forensic detection of cyborg propaganda specifically requires. What is needed is not a new legal instrument but a mandatory minimum scope definition within the existing Article 40 framework, one that explicitly encompasses network-level coordination data rather than leaving scope to platform discretion or DSC interpretation.

***Proposal 3: International Risk and Security Standards for Amplification (Macro Level)***

Third, international governance bodies should establish risk and security standards that require platforms to assess and mitigate the amplification of synthetically coordinated content, with financial penalties for recommendation algorithms that blindly prioritize engagement metrics regardless of coordination signals. This proposal adapts existing systemic risk assessment obligations, which already require very large online platforms to identify and mitigate risks to public discourse and electoral integrity (European Commission, 2024; Wolters & Zuiderveen Borgesius, 2025). The implementation challenge is metric design, specifically defining what constitutes a coordination signal for enforcement purposes without catching legitimate viral content. A staged approach, beginning with mandatory risk assessment disclosure before moving to penalty-linked thresholds, would allow regulators and platforms to develop workable definitions collaboratively. Risk assessment obligations should additionally encompass data poisoning as a second-order amplification harm. As synthetically coordinated content is indexed and incorporated into future AI training corpora, the corpus-level distortions it introduces compound the direct persuasion effects, skewing how downstream models represent political consensus. While a full governance framework for training data provenance lies beyond the scope of this paper, risk assessment mandates should at minimum require platforms to report on the proportion of high-engagement political content identified as synthetically coordinated, creating an evidentiary baseline for future regulatory action on training data integrity.

### *Micro-Level Interventions: Platform Design and Participatory Governance*

At the micro level, policy must address the user experience and public participation in AI governance. Because cyborg propaganda relies on minimizing the cognitive friction required to participate in a coordinated campaign, platforms should explore governance modes focused on implementation and design. This involves shifting content moderation strategies away from post-hoc deletion of speech, which risks infringing on the free speech rights of verified citizens, toward structural friction. Platforms could limit the rate at which users can authorize and disseminate API-generated text, forcing a degree of deliberative engagement before ratification. Such rate-limiting interventions must, however, be calibrated carefully to avoid disproportionately burdening users who rely on AI-assisted drafting for accessibility reasons, including non-native speakers and individuals with literacy or communication difficulties, for whom the friction imposed may constitute a barrier to legitimate political participation rather than a check on coordinated manipulation.

More fundamentally, governing cyborg propaganda requires moving beyond participation in campaigns toward participation in governance itself. Current platform architectures offer citizens two roles: targets of content moderation, or subjects of terms-of-service agreements that bear no meaningful relationship to informed consent (Obar & Oeldorf-Hirsch, 2020). Neither constitutes genuine participation in the governance of systems shaping their political environment. A more substantive approach involves multiple, layered forms of public engagement.

At the most immediate operational layer, community-driven labeling, analogous to context note systems but extended to coordination pattern identification, can distribute the detection burden without requiring platform or state authority to make determinative judgments about speech. Such systems allow users to evaluate the authenticity of the overarching narrative and its organizational origins, rather than solely judging the individual poster. This approach is more democratically defensible than algorithmic suppression, but its effectiveness depends on whether the labeling community is itself representative and insulated from the campaigns it is designed to flag, a non-trivial condition when coordination hubs can deploy participants to flood labeling systems and manufacture false or suppressive signals at scale (Draws et al., 2022; Pretus et al., 2024).

At a deeper structural layer, participatory governance requires incorporating affected publics into the design of detection and friction systems themselves, not merely their operation. This includes deliberative processes for setting rate-limiting thresholds, co-designing transparency labels, and auditing algorithmic curation (i.e., tasks currently delegated entirely to platforms). The challenge, as Sloane et al. (2022) argue, is that meaningful participation is not a design fix, but a structural commitment to incorporating affected communities into the governance process itself. Nominal participation mechanisms, consultations without binding input, advisory panels without agenda-setting power, risk legitimating platform authority rather than constraining it (Arnstein, 1969; Katzenbach & Ulbricht, 2019). Distinguishing genuine participatory governance from its simulation is especially critical here because the same infrastructure that manufactures political consensus can be turned toward manufacturing the appearance of public consent to governance arrangements.

Embedding participatory commitments into the architecture of platform content policies, rather than treating them as supplementary transparency measures, is therefore essential to building the public trust that disclosure-based interventions will otherwise fail to produce. It is also, as the comparative analysis below is intended to clarify, the governance dimension most sensitive to domestic political context. Participatory mechanisms that are credible in high-trust democratic environments may function as instruments of co-optation or surveillance in others.

The viability of each proposal varies significantly across regulatory contexts, and these differences are themselves analytically instructive (see Table 2). In the European Union, the DSA and AI Act provide meaningful legislative hooks for proposals two and three, and the EU's established tradition of platform regulation gives enforcement bodies legitimate authority (European Commission, 2024; Wolters & Zuiderveen Borgesius, 2025). The primary constraint is implementation speed. DSA enforcement remains fragmented across national Digital Services Coordinators, and the AI Act's transparency obligations for general-purpose AI systems are only now entering full force. In the United States, the First Amendment places categorical limits on content-level intervention, making the PAC registration model, which targets funding and organizational transparency rather than speech itself, the most constitutionally viable pathway. However, Federal Election Commission gridlock and the current regulatory posture toward platform self-governance make near-term federal action unlikely. Therefore, state-level electoral disclosure law may prove a more agile vehicle.

Critically, in non-democratic contexts, the governance calculus is reversed. The same coordination infrastructure that democracies seek to regulate may be actively deployed by governments to manufacture displays of popular loyalty and suppress dissent. This points to a structural asymmetry in international governance. The states most likely to adopt strong regulatory frameworks are also those most constrained by rule-of-law commitments, while the actors most likely to weaponize cyborg propaganda face no comparable domestic accountability. This asymmetry underscores the necessity of the international risk standards proposed above as the primary mechanism through which cross-border accountability for platform amplification can be established independent of national political will.

The comparative analysis presented here necessarily simplifies a more varied regulatory landscape. The EU/US/non-democratic tripartition captures the primary governance logics but leaves significant middle ground unaddressed. Canada, for instance, combines U.S.-adjacent platform governance norms with a functioning electoral finance disclosure regime that may be more immediately extensible to coordination hub registration than either the EU or federal U.S. frameworks. The United Kingdom's Online Safety Act and the emerging regulatory posture of Ofcom represent a third democratic model that assigns systemic risk obligations to platforms without the EU's multi-actor enforcement architecture. India and Brazil, as large democracies with active digital influence environments and distinct intermediary liability regimes, complicate any simple democratic/non-democratic dichotomy. These cases are not developed here, and the viability assessments in Tables 1 and 2 should be read as analytical anchors rather than exhaustive comparisons. A subsequent empirical study mapping cyborg propaganda governance across a broader set of national legal systems would substantially advance the comparative research agenda this paper opens.

**Table 1**
*Multilevel policy responses to cyborg propaganda*

| Level of analysis | Target entity | Legal basis | Key policy proposal | Key implementation challenge |
|---|---|---|---|---|
| Micro | • Users and platforms | • DSA systemic risk obligations (Art. 34)<br>• Platform terms of service | • Introduce rate-limiting friction for API-generated content authorization.<br>• Implement community-driven context labeling to identify coordinated campaigns. | • Defining thresholds that slow automation without suppressing legitimate AI-assisted speech.<br>• Accessibility risks for non-native speakers. |
| Meso | • Coordination hubs<br>• Electoral commissions | • Electoral finance disclosure law (FEC/EU Electoral Act)<br>• GDPR and equivalent data protection frameworks | • Classify coordination hubs as undisclosed political action committees.<br>• Apply data protection frameworks to network harvesting of third-party contact lists. | • Defining 'pre-authoring' vs. facilitating distribution.<br>• Jurisdictional variation in PAC-equivalent definitions across national legal systems. |
| Macro | • International regulators<br>• Large online platforms<br>• DSA Digital Services Coordinators | • DSA Art. 40 (researcher data access)<br>• DSA Art. 34 (systemic risk assessment)<br>• IGF coordination mechanisms | • Mandate researcher access to behavioral and network data for auditing synchronized cascades.<br>• Establish standards penalizing engagement-driven amplification of synthetic content.<br>• Require platforms to report synthetically coordinated high-engagement content, creating an evidentiary baseline for regulatory action on training data integrity. | • Defining mandatory minimum data scope; necessity and proportionality requirements risk being used to exclude network-level coordination data.<br>• Defining 'coordination signals' for enforcement without catching legitimate viral content. |

**Table 2**

*Comparative viability of governance proposals across regulatory contexts. Near-term = implementable within current legal frameworks with political will; medium-term = requires new legislation or significant regulatory reinterpretation; inverted = the governance logic is reversed, making domestic regulation inapplicable.*

| | European Union | United States | Non-democratic contexts |
|---|---|---|---|
| **Proposal 1: Classify coordination hubs as political action committees (meso level)** | | | |
| Viability | Medium-term | Near-term (state level) | Inverted |
| Legal hook | National electoral finance law; EU Electoral Act (limited scope) | FEC disclosure rules; state campaign finance law | No equivalent; same infrastructure deployed by state actors |
| Key constraint | No EU-level electoral finance regime; fragmented national definitions of "political expenditure" create regulatory arbitrage across member states | First Amendment limits content-level intervention; FEC gridlock blocks federal action; state-level disclosure law is the more agile near-term vehicle | Governance logic inverts entirely: the coordination infrastructure democracies seek to regulate is actively deployed by governments to manufacture displays of loyalty |
| Primary actor | National electoral commissions; European Parliament (longer term) | State electoral commissions; FEC (if gridlock resolved) | International bodies only (no domestic accountability mechanism) |
| **Proposal 2: Mandate independent researcher access to behavioral and network data (macro level)** | | | |
| Viability | Near-term | Low near-term | Inverted |

| | European Union | United States | Non-democratic contexts |
|---|---|---|---|
| Legal hook | DSA Art. 40 (researcher data access); existing mechanism, requires broader implementation | No direct equivalent; platform self-governance norm; voluntary agreements only | Platforms operate under state censorship; independent research structurally suppressed |
| Key constraint | DSA Art. 40 mechanism is legally established but implemented inconsistently across Digital Services Coordinators; scope of access remains contested | No statutory mandate; Congress has not acted; platforms treat data access as proprietary; academic-platform agreements are voluntary and easily withdrawn | State actors hostile to independent auditing; platforms in these markets operate under data-sharing obligations to governments, not researchers |
| Primary actor | EU Digital Services Coordinators; European Centre for Algorithmic Transparency | Congress (blocked); platforms voluntarily; bilateral agreements | International governance bodies; cross-border enforcement only |
| **Proposal 3: Risk and security standards penalizing amplification of synthetically coordinated content (macro level)** | | | |
| Viability | Near-term | Low near-term | Cross-border only |
| Legal hook | DSA Art. 34 (systemic risk assessment); AI Act obligations for very large platforms | FTC Act (unfair/deceptive practices, limited); no platform risk assessment mandate | International risk standards (IGF coordination mechanisms); no domestic hook |
| Key constraint | Enforcement lag between legal obligation and penalty-linked thresholds; defining "coordination signal" without catching legitimate viral content requires collaborative metric development | Section 230 immunizes platforms from third-party content liability; deregulatory posture limits FTC action; First Amendment restricts mandating algorithmic changes based on content | States most likely to weaponize cyborg propaganda face no comparable domestic accountability; international standards are the primary mechanism for cross-border |

| | European Union | United States | Non-democratic contexts |
|---|---|---|---|
| | | | enforcement independent of national political will |
| Primary actor | EU Digital Services Coordinators; AI Office; very large online platforms | FTC (limited authority); state attorneys general; voluntary platform commitment | IGF; bilateral platform governance agreements; UN GGE processes |

*Note.* Near-term: Implementable within current frameworks; Medium-term: Requires new legislation or reinterpretation; Low near-term: Structural barriers block near-term action; Inverted Governance: logic reversed; domestic regulation inapplicable.

## Implications for Future Research and Regulatory Implementation

The governance proposals advanced above are designed to operate within existing legal frameworks and do not await further empirical confirmation to be pursued by regulators. Their long-term durability and iterative refinement, however, depend on developing three streams of empirical and behavioral research that are currently underdeveloped. Each stream directly informs a specific aspect of the regulatory framework: forensic detection tools underpin the definitional thresholds that the PAC classification and risk standards require; behavioral research on user participation informs the calibration of micro-level friction and labeling interventions; and persuasion research on disclosure effects determines whether labeling obligations function as substantive governance instruments or as compliance formalities.

First, governance depends on forensic methods that target the infrastructure of coordination rather than the output of individual accounts. Traditional detection, based on syntax repetition, bursty timing, or shallow account histories, is defeated by cyborg propaganda's use of authentic accounts generating stylistically unique content. Effective regulatory enforcement therefore requires network-level forensics: Coordination indices that quantify hyper-synchronicity in posting times, thematic clustering that defies natural diffusion patterns, and analyses of account centrality and 'flipping' behavior between human and automated classifications (c.f. Ng et al., 2024; Schroeder et al., 2026). Since coordination hubs are often commercial products sold openly, supply-chain forensics, including passive DNS tracing and audit studies of volunteer recruitment interfaces, offer an additional detection pathway that does not depend on platform data access. Mapping these vulnerabilities provides the empirical foundation for the definitional thresholds that the PAC classification and risk standard proposals require.

Second, implementing the micro-level governance proposals depends on understanding the psychological conditions under which users participate in cyborg campaigns. Whether users experience participation primarily as strategic agency or as cognitive delegation (Köbis et al., 2025) has direct implications for the design of rate-limiting friction and community-labeling systems. If participation is predominantly driven by collective efficacy (van Zomeren et al., 2008), friction-based interventions may backfire by reducing the perceived impact of legitimate coordination. Longitudinal studies of AI-mediated political production, including the effects of posting AI-generated arguments on users' own political attitudes (see Jakesch et al., 2023), will be essential to calibrating and iteratively refining platform design interventions that protect legitimate speech while disrupting coordinated synthetic dissemination.

Third, assessing whether disclosure labeling reduces the persuasive impact of cyborg propaganda, or instead induces generalized skepticism toward all political speech, will determine how disclosure labeling obligations are best designed and calibrated. Inoculation-based experimental designs (Lewandowsky & van der Linden, 2021; Roozenbeek et al., 2022) offer a methodological template, but must be adapted to test the specific dynamics of 'relational shielding': the hypothesis that AI-generated messages transmitted via genuine social ties retain persuasive force even when recipients are aware of their synthetic origin. The answer will shape whether labeling mandates can be designed as effective governance instruments or risk becoming compliance formalities, an empirical question that should inform regulatory design rather than delay it.

## Conclusion

The rise of coordinated, AI-driven partisanship suggests that the public square has become an algorithmic battlefield. If we view cyborg propaganda as merely involving puppets, we risk dismissing the genuine frustrations that may drive individuals to join these automated activism platforms in the first place. Yet if we embrace it uncritically, we risk accepting a reality in which public opinion is merely another commodity manufactured on an assembly line, and in which the deciding factor in democratic contests is not whose ideas are better, but whose coordination app is more efficient.

The governance challenge this creates is distinct from prior problems of disinformation and bot manipulation, because cyborg propaganda is specifically engineered to operate within existing legal categories rather than in defiance of them. It cannot be addressed by extending frameworks built on the human/bot binary, because it systematically exploits that binary as a legal shield. Effective regulation must therefore be structural: targeting the commercial infrastructure that industrializes political coordination rather than the content it produces or the individual users who disseminate it.

To that end, this paper has proposed three concrete and legally grounded responses. Classifying commercial coordination hubs as political action committees shifts the disclosure burden to the operators and funders of cyborg campaigns without infringing on the speech rights of participants. Mandating researcher access to behavioral and network data, through mechanisms already partly established by existing policy, provides the forensic foundation that enforcement requires. And establishing international risk standards that penalize platforms for amplifying synthetically coordinated content addresses the one governance dimension that national frameworks cannot reach alone.

The comparative analysis reveals a structural asymmetry that policymakers must confront directly. The states most capable of enacting these reforms are simultaneously the most constrained by rule-of-law commitments, while the actors most likely to weaponize cyborg propaganda at scale face no comparable domestic accountability. This asymmetry makes international coordination not merely desirable but necessary. The challenge for the coming decade is not only to develop the behavioral and forensic models that detection requires, but to build the international governance architecture that makes detection actionable, distinguishing the roar of the crowd from the hum of the machine, and ensuring that distinction carries legal consequence.